\documentclass[preprint,eqsecnum,aps,prd,nofootinbib]{revtex4}

\usepackage{subfigure}%
\usepackage{graphicx}%
\usepackage{enumerate}%
\usepackage{amsmath}%
\usepackage{amsfonts}
\usepackage{verbatim}

\def\be{\begin{equation}}
\def\bea{\begin{eqnarray}}
\def\ee{\end{equation}}
\def\eea{\end{eqnarray}}
\newcommand{\beq}{\begin{equation}}
\newcommand{\eeq}{\end{equation}}
\newcommand{\beqa}{\begin{eqnarray}}
\newcommand{\eeqa}{\end{eqnarray}}
\newcommand{\beqar}{\begin{eqnarray*}}
\newcommand{\eeqar}{\end{eqnarray*}}

\newcommand{\al}{\alpha}

\newcommand{\n}{\nabla}
\newcommand{\s}{\sigma}

\def\openone{\leavevmode\hbox{\small1\kern-3.3pt\normalsize1}}%

\def\roughly#1{\raise.3ex\hbox{$#1$\kern-.75em\lower1ex\hbox{$\sim$}}}

\begin{document}
\title{The Physical Process First Law for Bifurcate Killing Horizons}

\author{Aaron J. Amsel\footnote{\tt amsel@physics.ucsb.edu}, Donald Marolf\footnote{\tt marolf@physics.ucsb.edu}, and Amitabh
Virmani\footnote{\tt virmani@physics.ucsb.edu}}

\affiliation{Department of Physics \\ University of California,
Santa Barbara \\ Santa Barbara, CA 93106, USA }

\begin{abstract}
The physical process version of the first law for black holes states
that the passage of energy and angular momentum through the horizon
results in a change in area $\frac{\kappa}{8 \pi} \Delta A = \Delta
E - \Omega \Delta J$, so long as this passage is quasi-stationary.
A similar physical process first law can be derived for any
bifurcate Killing horizon in any spacetime dimension $d \ge 3$ using
much the same argument.   However, to make this law non-trivial, one
must show that sufficiently quasi-stationary processes do in fact
occur.  In particular, one must show that processes exist for which
the shear and expansion remain small, and in which no new generators
are added to the horizon. Thorne, MacDonald, and Price considered
related issues when an object falls across a $d=4$ black hole
horizon.  By generalizing their argument to arbitrary $d \ge 3$ and
to any bifurcate Killing horizon, we derive a condition under which
these effects are controlled and the first law applies.   In
particular, by providing a non-trivial first law for Rindler
horizons, our work completes the parallel between the mechanics of
such horizons and those of black holes for $d \ge 3$.  We also
comment on the situation for $d=2$.
\end{abstract}

\date{August 2007}

 \maketitle

\tableofcontents

\section{Introduction}
\label{intro}

The analogy between the laws of black hole mechanics and
thermodynamics has been deeply ingrained in theoretical physics for
some time.  In their contribution to Bekenstein's festschrift,
Jacobson and Parentani \cite{JP} emphasized that these laws also
hold for much more general horizons, and in particular for what they
call asymptotic Rindler horizons.  Such horizons are the
boundary of the past of an event at future null infinity ($I^+$) in an asymptotically flat
spacetime.  In many cases such horizons result from small
perturbations of a Rindler horizon in flat spacetime.

Although \cite{JP} emphasizes the generality of horizon
thermodynamics, in their discussion the ``physical process version''
of the first law appears to be an exception.  This law describes the
dynamical change in horizon area in response to a flux of stress
energy through the generators.  As was first demonstrated by Hartle
and Hawking \cite{HH} (see also \cite{CarterReview} for a review),
for black holes this response can be written in the form
\begin{equation}
\label{first} \Delta E = \frac{\kappa}{8 \pi } \Delta A + \Omega
\Delta J.
\end{equation}
The form of this expression motivated \cite{Wald} to dub this result
the ``physical process version of the first law.''

Despite the issues raised in \cite{JP}, this note clarifies the
situation and demonstrates that the physical process first law holds
non-trivially for a general bifurcate Killing horizon in $d \ge 3$
spacetime dimensions.  In particular, it holds non-trivially for $d
\ge 3$ Rindler horizons.

As suggested by \cite{TJ1}, the first step in the argument is a
straightforward generalization of the one for black holes.  For
sufficiently quasi-stationary processes, one may follow
\cite{CarterReview,Wald} in using the Raychaudhuri equation for null
geodesic congruences to derive (\ref{first}).  By `sufficiently
quasi-stationary,' we mean that $i$) the expansion and shear  along
each generator are weak enough to neglect second order terms and
$ii$) no new generators are added to the horizon.  We give this
argument in section \ref{physical} below, elaborating on subtle
points.

The main issue raised in \cite{JP}  was whether sufficiently
quasi-stationary processes exist in the context of Rindler horizons.
Of particular concern was a result of Thorne, MacDonald, and Price
\cite{membrane} which states that, in $d=4$ spacetime dimensions,
the absorption of an object of mass $m$ and radius $r$ by a black
hole of mass $M$ will result in the formation of caustics when $r
\ll \sqrt{mM}$ in units with $G=1$ (while caustics fail to form when
$r \gg \sqrt{mM}$).  Considering a Rindler horizon as the $M
\rightarrow \infty$ limit of a black hole would therefore seem to
indicate that caustics always form when an object with any $m, r$
passes through a Rindler horizon.  But the formation of caustics
causes two problems:  $i$) in the region near caustics the expansion
becomes large and $ii$) caustics generally signal the addition of
generators to the horizon.  Thus, the above argument for the
physical process first law would not apply if one was forced to
consider horizon generators in a region with caustics.

We clarify the issues surrounding caustic formation in section
\ref{caustics} for the case $d \ge 4$.  There are two subtleties.
First, we emphasize that the derivation of the physical process
first law traces generators of the horizon only back to the
unperturbed past horizon. Thus, it is only to the future of the past
horizon that we need to avoid caustics.  In the $d=4$ black hole
context, it is precisely for this regime that the threshold $r \sim
\sqrt{mM}$  of \cite{membrane} determines whether caustics are
formed.  However, this expression for the threshold is only valid
when the object can be thought of as having begun its fall from rest
an infinite distance away from the black hole -- a condition which
does not admit a smooth limit to the case of Rindler horizons.  In
section \ref{caustics}, we show that (for $d=4$) a more local
characterization of when caustics form is given by the condition  $r
\sim \sqrt{E_\chi/\kappa}$.  Here $E_\chi$ and $\kappa$ are
respectively the Killing energy of the incident object and the
surface gravity defined by the Killing field $\chi$.  As usual, the
ratio $E_\chi/\kappa$ is independent of the choice of normalization
for the Killing field. We show that this condition applies to the
passage of a weakly self-gravitating object through any bifurcate
Killing horizon so long as $r$ is much smaller than the curvature
scale of the unperturbed horizon geometry. (Strongly
self-gravitating objects, such as black holes, generically lead to
the formation of caustics when they pass through a null congruence.)
We also show that the generalization of the above condition  to
arbitrary spacetime dimension $d \geq 4$ is
\begin{equation}
 \label{cc}
 r \sim \left(\frac{E_\chi}{\kappa}\right)^{\frac{1}{d-2}}.
\end{equation}
In section \ref{LowD}, we show that this conclusion also holds for
$d=3$, but argue that there is no general analogue for $d=2$. In
fact, for $d=2$ Rindler horizons in translation-invariant theories,
we argue that either the first law fails to hold or that it holds
only vacuously, in the sense that quasi-stationary processes do not
arise even in a limiting sense from physical processes. This last
point seems to resolve a certain tension \cite{GKS,ADV} associated
with particular viewpoints on gravitational entropy.  Finally,
section \ref{disc} provides some physical interpretation for
condition (\ref{cc}).

\section{The First Law far from Caustics}
\label{physical}

We now state the argument for the physical process version of the
first law following \cite{CarterReview,Wald}.  The argument is
essentially as outlined in \cite{TJ1}, but for completeness we give
the argument in its entirety.  The derivation is based on the
Raychaudhuri equation for null geodesic congruences, and on the
corresponding equations for the shear and twist. We consider a
bifurcate Killing horizon, so that the twist vanishes. We allow any
such horizon in $d \ge 3$ spacetime dimensions, but require that the
process be quasi-stationary.  By this we mean specifically that,  at
least in the region to the future of the unperturbed past horizon,
$i$) the expansion and shear along each generator remain weak enough
to neglect second order terms and $ii$) no new generators are added
to the horizon. Conditions under which these assumptions are
justified will be discussed in section \ref{caustics}.

It is convenient to parametrize the geodesics in terms of the
Killing parameter $v$ associated with some horizon-generating
Killing field $\chi$.  With this understanding, the `focusing
equation' for the expansion becomes
 \be
  \frac{d \hat \theta }{d v} = \kappa \hat \theta - \frac{\hat \theta^2}{d-2}
   - \hat \sigma_{\mu \nu} \hat \sigma^{\mu \nu}
    - R_{\lambda \sigma} \chi^{\lambda} \chi^{\s},
 \label{focus} \ee
where $\kappa$ is the surface gravity defined by $\chi^\sigma
\nabla_\sigma \chi^\mu = \kappa \chi^\mu$ and the hats ($ \ \hat{} \
$) on the expansion and shear ($\hat \theta, \hat \sigma_{\mu \nu}$)
remind the reader that these quantities have been defined using the
Killing parameter $v$ (as opposed to the more usual affine parameter
$\lambda$).   Note that we have not fixed the normalization of
$\chi$; our final results will be independent of this normalization.
Equation (\ref{focus}) gives the standard result for $d=4$ and is
derived for the general case $d \ge 3$  in appendix \ref{rayeqs},
along with the corresponding equations for shear and twist.

As stated above, we assume that the expansion and shear are weak
enough that we may truncate (\ref{focus}) to linear order:
 \be - \frac{d \hat \theta}{dv}  + \kappa
\hat \theta = S(v), \label{source} \ee
 where we may use the Einstein equations to write the source as the
non-gravitational energy flux through the horizon
 \be
  S(v) = 8 \pi T_{\lambda \sigma} \chi^{\lambda} \chi^{\s}.
   \ee

As we wish to consider sources associated with brief departures from
equilibrium, we shall assume that $S(v)$ vanishes rapidly as $v
\rightarrow \pm \infty$, and that the expansion and the shear tend
to zero in the final configuration.  We may therefore solve
(\ref{source}) using an advanced Green's function: \be \hat
\theta(v) = \int_{v}^{\infty} e^{\kappa (v - v')} S(v') dv'.
\label{integral} \ee

Now, recall that the expansion of a null congruence measures the
fractional change in the area of a bundle of null generators
over a finite range of Killing time,
 \be
  \Delta A = \int_{\mathcal{B}} \hat \theta \: dA \: d v \,,
   \ee
where $\mathcal{B}$ is the piece of the horizon generated by the
bundle of null generators over the given range of Killing time. For
weak perturbations, the fractional change in the area is simply the
integral of the expansion over the Killing time.  The asymptotic
change in area $d(\Delta A)$ along a given generator  of initial
area $dA$ is then
 \be
  \frac{d(\Delta A)}{dA} = \int_{-\infty}^{\infty} \hat \theta \: dv = \int_{-\infty}^{\infty}  dv \int_{v}^{\infty} dv'
e^{\kappa (v-v')} S(v').
 \label{noncausal}
    \ee
Changing  the order of integration and integrating over $v$ one
finds
 \be
  \frac{d(\Delta A)}{dA}= \frac{1}{\kappa} \int_{-\infty}^{\infty}
dv'  S(v') = \frac{8 \pi }{\kappa} \int_{-\infty}^{\infty} dv \
T_{\mu \nu} \chi^{\mu} \chi^{\nu} .
 \label{twoterms}
  \ee
Since the integral of $T_{\mu \nu} \chi^{\mu} \chi^{\nu} $ over the
horizon gives the flux through the horizon of Killing energy
$E_\chi$ associated with $\chi$, we have derived the first law:
 \be
 \label{1LawnoJ}
 \frac{\kappa \Delta A}{8 \pi } = \Delta E_\chi.
 \ee
The more general version of the first law with angular momentum flux
follows immediately in the case where one uses different Killing
fields $t^\mu,\phi^\mu$ to define energy $E$ and angular momentum
$J$ and where $\chi^\mu = t^\mu + \Omega \phi^\mu$.  In this case
(\ref{1LawnoJ}) becomes
 \be
 \label{1Law}
 \frac{\kappa \Delta A}{8 \pi } = \Delta E - \Omega \Delta J.
 \ee

 Let us comment briefly
on the physical interpretation of this law, and in particular on the
left-hand side.  Recall that we computed $\Delta A$ by integrating
the expansion over $v \in (-\infty, \infty)$.  Since we used
advanced boundary conditions, it is clear that $v = \infty$ is the
asymptotic future.  On the unperturbed future horizon,  $v =
-\infty$ was the bifurcation surface where the future and past
horizons intersect. In fact, even with the perturbation we may
repeat the above derivation replacing the past limit of integration
$v = -\infty$ with the surface where our generators intersect the
(unperturbed) past horizon.  The point is that, since we take second
order terms to be small, $v = -\infty$ can differ from this surface
only by at most a first-order error term. But since the expansion
(i.e., the integrand) is also of first order, this means that
integrating back to the past horizon changes (\ref{twoterms}) only
by a second order term.  Thus, the correction is negligible.

The advantage of tracing the generators back to a cross-section of
the unperturbed past horizon is that, since the unperturbed horizon
is at equilibrium, the area of any such cross section is just the
area of the unperturbed horizon.  Thus, as desired, the left-hand
side of (\ref{1Law}) represents the difference between the area of
the perturbed horizon in the asymptotic future and the area of the
unperturbed horizon.

Now, in practice, there is typically even more flexibility in
choosing the past limit of integration.  Note that all of the
integrals in the derivation converge, and that the characteristic
response time associated with the solution (\ref{integral}) is of
order $\kappa^{-1}$.  Thus, if the perturbation is well-localized in
time, one may think of $\Delta A$ as describing the change in area
between times $v_i, v_f$ which precede and follow the perturbation
by any interval significantly greater than $\kappa^{-1}$.  With this
interpretation, it is clear that the first law in fact applies to
many horizons which only approximate a bifurcate Killing horizon.
For example, it applies not only to strict Rindler horizons
associated with an exact boost symmetry, but also to the asymptotic
Rindler horizons of \cite{JP}.  (See also \cite{TJ1}, where they
were called ``partial horizons.'')

\section{Characterizing the Formation of Caustics}
\label{caustics}

In section \ref{physical} above we considered the physical process
first law for quasi-stationary processes.  By `quasi-stationary' we
mean processes in which, at least to the future of the unperturbed
past horizon, $\hat \theta, \hat \sigma_{\mu \nu}$ remain weak
enough to ignore all second order effects and to avoid the addition
of new generators to the horizon.  One expects that sufficiently
weak perturbations are quasi-stationary in this sense.   However, it
is important to check that there do indeed exist perturbations for
which this is the case.

Regarding the expansion, we see from (\ref{focus}) that the $\hat
\theta^2$ term becomes relevant when $\hat \theta \sim
\kappa/(d-2)$.  To understand the effects of this term, let us
consider the solution to (\ref{focus}) in a region where $\hat
\sigma^2= 0 = S(v)$.  As noted in \cite{membrane}, the desired
solution is then
\begin{equation}
\bar \theta(v) = \frac{1}{1+\left( \bar
\theta_0^{-1}-1\right)e^{\kappa(v_0-v)}} \,,
\end{equation}
where $\bar \theta = \frac{\hat \theta}{(d-2) \kappa}$ and $\bar
\theta(v_0) = \bar \theta_0$.  If $\bar \theta_0 < 1$, then $\bar
\theta$ decreases toward zero as $v$ decreases into the past.  If,
however, $\bar \theta_0 > 1$, then $\bar \theta$ increases into the
past and diverges at some finite time.  Therefore, if the horizon is
perturbed strongly enough to cause $\bar \theta \gtrsim 1$ (i.e.,
$\hat \theta \gtrsim  \kappa/(d-2)$) at any $v = v_0$, then the
focusing equation implies that a caustic developed at  some $v <
v_0$.  Thus, the requirement that non-linear terms can be ignored is
essentially the requirement that no caustics form.  This is also a
necessary condition to avoid the addition of new generators.

Below, we generalize the discussion of \cite{membrane, Suen} to show
that when a small, weakly self-gravitating object  passes through an
arbitrary bifurcate Killing horizon, the condition (\ref{cc}) sets
the threshold for caustic formation to the future of the unperturbed
past horizon.  By ``small and weakly self-gravitating,'' we mean
that the radius $r$ satisfies $m^{1/{d-3}} \ll r \ll \ell$, where
$\ell$ is the background curvature scale near the horizon.   We
consider here the case $d \ge 4$; lower dimensions will be discussed
in section \ref{LowD}.

We noted above that the evolution of the expansion $\hat \theta$ is
controlled by the focusing equation (\ref{focus}) of section
\ref{physical}.  We will also require the corresponding `tidal-force
equation' which governs evolution of the shear $\hat \sigma_{\mu
\nu}$ along a congruence with vanishing twist. This equation is
derived in appendix \ref{rayeqs} and takes the form
 \be
 \frac{d\hat \sigma_{\mu \nu}}{d v} = \left( \kappa  - \frac{2 \hat
 \theta}{d-2}\right) \hat \sigma_{\mu \nu}- \hat \s_{\mu \s}\hat
 \s^{\s}{}_{\nu}+ \frac{ \hat \s^2}{d-2}Q_{\mu \nu}  + \left( 2 \hat
 \s_{\mu \s} + \frac{2 \hat \theta}{d-2} Q_{\mu \s}\right) \hat
 \s^{\s}{}_{\nu}-  \mathcal{E}_{\mu \nu}.
  \label{tidal}
   \ee
Here, for each $v$, the tensor $Q_{\mu \nu}$ is the projector onto
the spacelike cut of the horizon naturally associated with constant
Killing time $v$ as explained in appendix \ref{rayeqs}.  The final
source term $\mathcal{E}_{\mu \nu} := Q^{\al}{}_{\mu}
Q^{\beta}{}_{\nu} C_{\alpha \lambda \beta \s} \chi^{\lambda}
\chi^{\s} $ is the electric part of the Weyl tensor.

Note that if the energy flux through the horizon is of first order
in a small dimensionless perturbation parameter $\epsilon$, then so
is $\mathcal{E}_{\mu \nu}$.  We now distinguish between generators
which intersect the matter and generators which do not.  For those
which do, the horizon perturbations $ \hat \theta$ and $\hat \s_{\mu
\nu}$ are again of order $\epsilon.$  However, for those which do
not, we see that the expansion $\hat {\theta}$ is only of second
order in $\epsilon$.  As a result, we will need to keep the $\hat
\sigma^2$ term in the focusing equation (\ref{focus}) below.  We
will, however, drop the term $\hat \theta^2$.

Truncating the tidal force equation to $\mathcal{O}(\epsilon)$ and
the focusing equation to $\mathcal{O}(\epsilon^2)$ yields
 \be -
\frac{d \hat \sigma_{\mu \nu}}{d v}
 + \kappa \hat \sigma_{\mu \nu} = \mathcal{E}_{\mu \nu},
  \ee
and
\be
 - \frac{d \hat \theta}{dv}  + \kappa \hat \theta = S(v) +
\hat \sigma^2 \,.
 \ee
As in section \ref{physical}, the desired solutions follow by
integrating the sources against an advanced Green's function:
 \be
  \label{weaksol1}
   \hat \sigma_{\mu
\nu}(v) = \int  \mathcal{E}_{\mu \nu}(v') \ e^{\kappa (v-v')}
\Theta(v'-v) dv' \ee \be \label{weaksol2} \hat \theta(v) = \int
\left( S(v')  + \hat \sigma^2(v') \right) e^{\kappa (v-v')}
\Theta(v'-v) dv'.
 \ee

We wish to apply this analysis to the situation in which an object
of mass $m$ falls freely through the horizon.  Following
\cite{membrane}, our strategy will be to describe this process as
the passage of the mass through a Rindler horizon in flat spacetime.
One might think that this requires the curvature of the spacetime to
be small.  Indeed, as stated above, we require the curvature scale
of the spacetime near the horizon to satisfy
 $ \ell \gg r$.  However, we
need make no further restrictions on $\ell$.  To see this, note
first that since $\chi$ generates an isometry the scale $\ell$ is
invariant under the diffeomorphism generated by $\chi$. This
diffeomorphism acts like a flat-space boost near the bifurcation
surface.  Thus, we may approximate any region near the bifurcation
surface as being in flat spacetime, so long as there is {\it some}
reference frame in which the size of this region is much less than
$\ell$ as measured by the corresponding locally inertial
coordinates.

One might ask if we can choose such a frame so that this small
region includes the event where the center of our mass falls across
the future horizon. But since this event is null-separated from the
bifurcation surface, it is clear that such a choice is always
possible!  We need only choose a sufficiently ``boosted'' frame in
which the coordinate separation of this point from the bifurcation
surface is small. Furthermore, since we assumed $r \ll \ell$ above,
it follows that every event where the horizon intersects our object
is in fact contained in the desired region.

Note that, having determined that a flat space analysis is valid in
this particular frame, we are free to apply an additional boost  to
transform this flat-space description to any other convenient frame.
Below, we will choose the frame in which the object is at rest, say
at $Z = z_0$, $x_i = 0$ in terms of the usual Minkowski coordinates
$(T, Z, x_i)$.

Let us first consider those geodesics through which the matter
energy flux is negligible.  In other words, we consider geodesics
that pass outside the object itself.  Since we assumed $r \gg
m^{1/(d-3)}$,  we may describe the object by a linearized solution
to the Einstein equations.   For simplicity, we take the object to
be spherically symmetric; the general case follows by linear
superposition.  The corresponding linearized metric takes the form
\begin{eqnarray}
ds^2 &=& -\left(1-\frac{c \, m}{\left(\sqrt{\rho^2 + (Z-z_0)^2}\right)^{d-3}} \right)\, dT^2 \nonumber \\
&& \qquad \qquad \quad+ \left(1+\frac{1}{d-3}\frac{c \,
m}{\left(\sqrt{\rho^2 + (Z-z_0)^2}\right)^{d-3}} \right)
\left(dZ^2+d\rho^2+\rho^2 d\Omega^2_{d-3} \right),
\end{eqnarray}
where  $\rho^2 = x_1^2+\ldots+x_{d-2}^2$ and the line element on the
$S^{d-3}$ is parametrized by angles $\phi_j$, with $j = 1, \ldots,
d-3$.  The constant $c$ is
\begin{equation}
c = \frac{16 \pi}{(d-2) \Omega_{d-2}} \,,
\end{equation}
where $\Omega_n$ is the volume of $S^n$. The diagonal transverse
components of the electric part of the Weyl tensor for this metric
are
\begin{equation}
\label{weyl1} \mathcal{E}_{\rho \rho} = -\frac{d-3}{g_{\phi_j
\phi_j}} \mathcal{E}_{\phi_j \phi_j} = - \frac{(d-3)(d-1) c \, m
\kappa^2}{2} \, \frac{\rho^2 T^2}{\left(\sqrt{\rho^2 +
(Z-z_0)^2}\right)^{d+1}} \, ,
\end{equation}
while all the components with mixed indices vanish, $i.e.,$ $E_{\rho
\phi_j} = 0 = E_{\phi_j \phi_k} (j \neq k)$.

We wish to evaluate (\ref{weyl1}) along the horizon $T-Z=0$ and to
express the result in terms of the advanced Killing time
 \be
 v =  \kappa^{-1} \ln \left( \frac{\kappa}{2} (T+Z) \right)
 \ee
used in (\ref{weaksol1}, \ref{weaksol2}).  It is useful to work in
terms of the shifted coordinate $\bar v = v - v_0$, where $v_0$ is
the value of $v$ where the object crosses the horizon.  Note that
along the horizon we have
\begin{equation}
T = z_0 e^{\kappa \bar v}\, .
\end{equation}
Using this relation in (\ref{weyl1}) and expanding the exponentials
for $\kappa \bar v \ll 1$ we have
\begin{equation}
\label{weyl2} \mathcal{E}_{\rho \rho} =  - \frac{(d-3)(d-1) c \, m
\kappa^2 z_0^2 }{2} \, \frac{\rho^2}{\left(\sqrt{\rho^2 + \left(z_0
\kappa \bar v \right)^2}\right)^{d+1}} \, .
\end{equation}
Our analysis will now simplify considerably if we approximate the
time dependence in this result by a delta function \cite{Suen},
 \be \label{weyl3}
  \mathcal{E}_{\rho \rho} = -(d-3) \frac{8 \pi m  \kappa
z_0 }{\Omega_{d-3}} \frac{\delta(\bar v)}{\rho^{d-2}} =
-\frac{d-3}{g_{\phi_j \phi_j}} \mathcal{E}_{\phi_j \phi_j} \,,
 \ee
which has the same time integral as (\ref{weyl2}).   Note that the
tidal forces then depend on $m$ and the object's trajectory only
through the quantity $m \kappa z_0$, which is just the Killing
energy $E_\chi = -p_\mu \chi^\mu$, where $p^\mu$ is the
four-momentum.  The replacement of (\ref{weyl2}) by (\ref{weyl3}) is
a good approximation when $\rho \ll z_0 = E_\chi/(m \kappa)$.  Let
us assume for the moment  that  $r \ll z_0 = E_\chi/(m \kappa)$, so
that $\rho \ll z_0 = E_\chi/(m \kappa)$ is the region of greatest
interest.  We will return to the more general case shortly.

Using (\ref{weyl3}) in our expression for the shear
(\ref{weaksol1}), we have
 \be
  \hat \sigma_{\rho \rho} =
-\frac{d-3}{g_{\phi_j \phi_j}} \hat \sigma_{\phi_j \phi_j} = -(d-3)
\frac{8 \pi E_\chi}{\Omega_{d-3}} \frac{e^{\kappa \bar v}
\Theta(-\bar v)}{\rho^{d-2}} \,.
 \ee
We then substitute this result into (\ref{weaksol2}) to find the
expansion,
 \be
  \hat \theta = \frac{d-2}{\kappa (d-3)} \left( \frac{8
\pi(d-3) E_\chi}{\Omega_{d-3}\rho^{d-2}} \right)^2 e^{\kappa \bar v}
(1- e^{ \kappa \bar v}) \Theta(-\bar v) \,.
 \ee
 The arguments given
above then imply that caustics will form along our geodesic if
 \be
 \left(\frac{\hat \theta}{(d-2) \kappa} \right)_{max}  \gtrsim 1\, ,
 \ee
 that is, if
  \be
  \label{OC1}
   r \lesssim  \left( \frac{4 \pi \sqrt{d-3}}{
\Omega_{d-3}}\frac{ E_\chi}{\kappa} \right)^{\frac{1}{d-2}} \, ,
 \ee
where we have set $\rho = r$, the radius of the object.  On the
other hand, no caustics will form along our geodesic if
  \be
  \label{OC}
   r \gg  \left( \frac{4 \pi \sqrt{d-3}}{
\Omega_{d-3}}\frac{ E_\chi}{\kappa} \right)^{\frac{1}{d-2}} \, .
 \ee
For $d=4$ and $\kappa^{-1} = 4 M$, this reduces exactly to the
result stated in \cite{membrane}.

The above results were derived using the approximation $r \ll z_0 =
E_\chi/(m\kappa)$ to replace (\ref{weyl2}) by (\ref{weyl3}).
However, it turns out that this need not be taken as an independent
assumption.  Let us first consider the case when (\ref{OC1}) holds.
Recall that we take the object to be weakly gravitating, so that
$r^{d-3} \gg m$. Combining this statement with (\ref{OC1}), we
arrive at $r \ll z_0 = E_\chi/(m\kappa)$, so that the replacement of
 (\ref{weyl2}) by (\ref{weyl3}) is justified.

On the other hand, let us consider the case when (\ref{OC}) holds.
For $r \gtrsim z_0$, the approximation of (\ref{weyl2}) by
(\ref{weyl3}) makes an error. However, one notes that the solution
(\ref{weaksol2}) has a tendency to ``forget'' about the source at
$v' > v$.  As a result, compressing the source to a delta-function
can only increase the effect on the expansion. Thus, the maximum
expansion resulting from (\ref{weyl2}) is strictly less than the
maximum resulting from (\ref{weyl3}).  It follows that condition
(\ref{OC}) forbids the formation of caustics along our geodesics
without further qualification.

We now consider those geodesics which do pass through the object.
For simplicity, we consider a homogeneous object of constant
Killing-energy density $E_\chi/r^{d-1}$.  In this case, we see from
(\ref{weyl3}) that the electric part of the Weyl tensor is smaller
inside the object than just outside.  But the Weyl tensor is the
only source for the shear.  Thus, when (\ref{OC}) holds, the shear
also remains small along all geodesics which intersect the matter.

It remains only to analyze the expansion, which is a linear
functional of $S(v) + \hat \sigma^2$.  Let us first note that for
the above source the contribution from $S(v)$ to $\hat
\theta/\kappa$ is bounded by a term of order $E_\chi/(\kappa
r^{d-2})$.  On the other hand, we saw above that the shear term
contributes a term of order $\left(E_\chi/(\kappa r^{d-2})
\right)^2$.  Adding the two such terms makes it clear that
conditions (\ref{OC1}), (\ref{OC}) again determine whether or not
caustics form in the region of interest.

\section{Lower Dimensions: $d=2,3$}
\label{LowD}

It is interesting to discuss the remaining cases of lower spacetime
dimension $d=2,3$. (There are no horizons for $d < 2$.) The analysis
for $d=3$ is quite similar to that above. The arguments of section
\ref{physical} already hold for $d=3$, and the main difference in
section \ref{caustics} is that both the Weyl tensor and the shear
vanish identically.  For those generators which pass through the
object, we have already seen that ignoring the shear changes the
threshold only by a coefficient of order one, so for $d=3$ we again
obtain $r \sim (E_\chi /\kappa)^{1/(d-2)} = E_\chi /\kappa$.    For
generators which pass outside the object, the expansion simply
remains zero. Locally, this part of the congruence is non-singular.
However, the left and right sides of the congruence will
nevertheless cross due to global effects.   So long as $r$ is much
smaller than any curvature scale in the unperturbed background, one
may use the conical deficit angle $\delta = 4m$ generated by a point
mass \cite{2+1} to check that the threshold for this to occur is
once again set by $r \sim E_\chi /\kappa= (E_\chi
/\kappa)^{1/(d-2)}$. Thus, our results extend to the case $d=3$.

 Let us now consider the case $d=2$.  Here the main complication
 is that the Einstein-Hilbert action becomes trivial.
Nevertheless, one may define a non-trivial theory either by studying
a scalar gravity theory (dilaton gravity) or by considering
compactifications of a higher-dimensional theory. The latter is, in
a sense, a special case of the former.  We proceed by considering
various examples.

 Let us first suppose that one simply compactifies
$n$-dimensional Einstein gravity on an $n-2$ torus.  One may again
then argue as in section \ref{physical} that a first law holds for
any quasi-stationary process.  However, since there is only one
uncompactified spatial direction,  the gravitational field induced
by any perturbation now tends to grow linearly with distance and can
even change the asymptotics of the spacetime.  There is thus no
analogue of the arguments in section \ref{caustics}.  In particular,
it seems likely that the passage of such an object would destroy any
asymptotic Rindler horizon.

Another sort of $d=2$ compactification arises when some method has
been used to `stabilize the moduli' at particular values  (see e.g.
\cite{fate,modstab} for reviews).
 In practice, this means incorporating various quantum
and/or stringy effects to create a potential for the size of the
compactified directions.  For a $d=2$ compactification, this
effectively creates a mass term for the gravitational degrees of
freedom and removes the linear growth of gravitational fields
described above. It thus stabilizes the boundary conditions.
However, it also removes us from the regime where the Einstein
equations alone can be used to study the response of the horizon.
In particular, since the volume of the compactified dimensions tends
to remain constant, the area of the horizon is not changed by the
passage of any object.  In fact, if all moduli are stabilized at
particular values, the horizon after the passage of the object will
be indistinguishable from the original horizon. Thus, no analogue of
the physical process first law can hold in this context.   For the
same reason, it is also clear that no stationary comparison version
of the first law can hold.   We therefore see no reason to assign a
finite entropy to asymptotic horizons in this context.

As a third example of $d = 2$ gravity, we consider dilaton gravity
associated with linear dilaton vacua.  In such theories the dilaton
runs from infinitely weak coupling on one side (say, at the right
infinity) to infinitely strong coupling at the other (left)
infinity.   Unlike the previous two examples, these theories can
admit black hole solutions.  In fact, all of the known 1+1 black
holes  \cite{WittenBH, CGHS} arise in this context.

Such linear dilaton boundary conditions turn out to be stable.
However, in contrast to the case where the dilaton modulus is
`stabilized' at a particular value, here the value of the dilaton
tends to change monotonically along horizons. As a result, in this
context, there {\it can} be a first law for both black hole and
asymptotic Rindler horizons.  However, since our methods do not
apply directly to this case, the first law must be verified by
another method (see e.g. \cite{CGHS}).

The above examples suggest that 1+1 gravity systems enjoy a first
law of horizon mechanics only when the boundary conditions break
asymptotic spatial translation symmetry.  This observation is
interesting in the context of \cite{GKS}, which predicted that there
would be no Poincar\'e-invariant 1+1 compactifications of consistent
quantum gravity theories (and of string theory in particular).  To
arrive at this claim, the authors assumed that, since $d=2$ Rindler
horizons have finite area, one can assign them finite
entropy\footnote{ More precisely, the authors of \cite{GKS} supposed
that $i$) compactifications of higher dimensional gravity with
stabilized moduli would lead to 1+1 horizons with finite entropy and
$ii$) the entropy of such horizons should agree with the von-Neumann
entropy of $\exp(-\beta H)$, where $\beta, H$ are respectively the
inverse temperature and the Killing energy operator associated with
the horizon. Ref. \cite{GKS} showed that ($i$) and ($ii$) conflict
with Poincar\'e symmetry, which led to the prediction stated above.
We note that, so long as one adds the assumption that $iii$) the
theory contains localized excitations, this conclusion continues to
hold if assumption ($ii$) above is replaced with the somewhat
different assumption that $ii'$) the entropy of such horizons counts
the total number of quantum states associated with the system behind
the horizon.  In fact, even spatial translations alone are enough to
cause a conflict with ($i$), ($ii'$) and ($iii$).  See \cite{dS} for
detailed comments on assumption ($ii$) in the original form.}. But
our observation above suggests that horizons in such theories do not
enjoy a first law.  Thus, we see no reason to assign them a finite
entropy. From this viewpoint, it is no surprise that \cite{ADV} did
in fact construct 1+1 Poincar\'e-invariant compactifications of
string theory which are free of massless moduli.

\section{Discussion}
\label{disc}

We have argued that the physical process version of the first law
holds non-trivially for any bifurcate Killing horizon in spacetime
dimensions $d \ge 3$.  We have also seen that it holds for $d \ge 3$
approximate Killing horizons such as general asymptotic Rindler
horizons.

 In addition to giving a straightforward derivation of the first
law for quasi-stationary processes, we generalized the arguments of
\cite{membrane,Suen} to processes where a homogeneous
weakly-gravitating object passes through any bifurcate Killing
horizon in any spacetime dimension $d \geq 3$.  In particular, we
showed that for $d \ge 3$ the condition
 \be
 \label{cc2}
 r^{d-2} \sim E_\chi/\kappa
 \ee
sets the threshold for the formation of caustics to the future of
the past horizon. When $r^{d-2} \gg E_\chi/\kappa$, no caustics form
in the region to the future of the unperturbed past horizon and the
process is indeed quasi-stationary. Thus, the first law applies. In
particular, for $d \ge 3$ our work completes the analogy between
black hole and asymptotic Rindler horizons outlined in
\cite{JP,TJ1}.

The condition (\ref{cc2}) should not be a surprise.  If the first
law holds, this threshold is
 \be
 \label{area}
 r^{d-2} \sim \Delta A.
 \ee
In other words, the first law is valid when the horizon area
(entropy) through which the object passes is much larger than the
change in area (entropy) induced by the object itself. From the
thermodynamic perspective, this is a natural definition of a
quasi-stationary process\footnote{We thank Ted Jacobson for
suggesting this interpretation of (\ref{cc}).}.

One may use a related perspective to briefly summarize the arguments
of section \ref{caustics} without going through the technical
details. The point is that one expects caustics to form first along
those geodesics which pass through the edge of the matter
distribution, where both the matter-energy density and the electric
part of the Weyl tensor act as sources for (\ref{focus}) and
(\ref{tidal}).  But as discussed in section \ref{physical}, dropping
non-linear terms in (\ref{focus}), (\ref{tidal}) implies that an
energy $\delta E_\chi$ passing through a bundle of generators with
area $\delta A$ causes a change in the area of that bundle given by
$\Delta (\delta A) = \delta E_\chi/\kappa$. So long as this is much
less than $\delta A$ itself one must be far from caustics, as
caustics arise when the area of a bundle of null geodesics shrinks
to zero. Thus a first-order treatment should be valid and the
physical process first law should hold.  Integrating this condition
over the matter distribution gives precisely (\ref{cc2}).

In contrast, all of the above results may fail for $d=2$.  In
particular, there can be no first law when higher-dimensional
gravity is compactified to $d=2$ in a manner that stabilizes all
moduli, or in the string theories of \cite{ADV}.  Such theories
effectively assign gravity a mass $m >0$, and turn off the
long-range interaction.  Thus, one might say that the long-distance
Newton's constant $G$ has been renormalized to zero.  This viewpoint
suggests that any horizon entropy is strictly infinite, and it is no
surprise that the first law becomes trivial.  On the other hand,
horizon entropy is often thought of as a short-distance phenomenon.
From this perspective one might not expect the mass $m$ to influence
the entropy, since $m$ may be much less than any fundamental scale
(such as the short-distance Planck Mass $m_{pl}$).  It would thus be
especially interesting to understand these effects from a
microscopic perspective.

\section*{Acknowledgements}
D.M. thanks Ted Jacobson for many interesting discussions on these
and related issues, and for detailed comments on an early draft of
the manuscript.   He also thanks Steve Giddings and Gary Horowitz
for a discussion of 1+1 compactifications.   This work was supported
in part by the National Science Foundation under Grant No
PHY05-55669, and by funds from the University of California.

\appendix

\section{Raychaudhuri Equations}
\label{rayeqs} This appendix reviews the derivation of the
Raychaudhuri equation for null geodesic congruences and the
associated equations for the shear and twist in general spacetime
dimension $d \ge 3$.  As in the main text, we consider a congruence
associated with a bifurcate Killing horizon\footnote{The
Raychaudhuri equations for expansion, shear, and twist in general
spacetime dimension were previously derived in \cite{OPP} using
affine parametrization instead of Killing parametrization.  It seems
that the expansion equation in general spacetime dimension has been
known for some time, e.g. \cite{CarterReview}.}.

Let $k^\mu$ be the affinely parametrized future-pointing null normal
generating the bifurcate Killing horizon $\mathcal{K}$.  On the
bifurcation surface, there is a second future-pointing null normal
 $l^\mu$, which one may think of as pointing in the opposite spatial direction to
$k^\mu$.  We choose $l^\mu k_\mu = -1$ and $k^\mu \nabla_\mu l^\nu =
0$ on the bifurcation surface, and define $l^\mu$ all along the
congruence by parallel transport along the geodesics.  Thus, the
above inner products hold at each point in the congruence.

For our purposes, it is more convenient to parametrize the null
geodesics so that their tangents are given by the Killing vector
field $\chi^\mu$, rather than $k^\mu$. This vector satisfies
\begin{equation}
\chi^\sigma \nabla_\sigma \chi^\mu = \kappa \chi^\mu,
\end{equation}
where $\kappa$ is the surface gravity of $\mathcal{K}$.  Similarly
we can now define a second null vector field $\hat l$ satisfying
$\hat l^\mu \chi_\mu = -1$ and $\chi^\mu \nabla_\mu \hat l^\nu =  -
\kappa \hat l^\nu$.  Now, following for example \cite{CarterReview},
we can define a projection tensor
\begin{equation}
Q_{\mu \nu} = g_{\mu \nu} + \chi_\mu \hat l_\nu+ \chi_\nu \hat l_\mu
\end{equation}
that projects onto the $(d-2)$-dimensional space spanned by the
deviation vectors orthogonal to both $\chi^\mu$ and $\hat l^\mu$.
This also coincides with the space tangent to the cut ${\cal C}$ of
the horizon obtained by Lie dragging the bifurcation surface along
the affine tangent field $k^\mu$.

We now introduce the distortion tensor
 \be
  \hat B_{\mu \nu} = Q^{\al}{}_{\mu} Q^{\beta}{}_{\nu} \n_{\beta} \chi_{\al} \, ,
 \ee
which satisfies
 \bea
 \chi^{\sigma} \n_{\sigma} \hat B_{\mu \nu} =
\kappa \hat B_{\mu \nu} - \hat B_{\mu}{}^{\s} \hat B_{\sigma \nu} -
Q^{\al}{}_{\mu} Q^{\beta}{}_{\nu}R_{\al \lambda \beta \s}
\chi^{\lambda} \chi^{\s}.
 \label{final}
  \eea

 The tensor $\hat B_{\mu \nu}$ can be decomposed into
expansion, shear, and twist as
\be
\hat B_{\mu \nu} = \frac{\hat \theta}{d-2} Q_{\mu \nu} + \hat
\s_{\mu \nu} + \hat \omega_{\mu\nu}, \label{ray1}
\ee
where $\hat \theta = Q^{\mu \nu} \hat B_{\mu \nu} $ , $\hat
\sigma_{\mu \nu} = \hat B_{(\mu \nu)} - \frac{\hat \theta}{d-2}
Q_{\mu \nu}$, and $\hat \omega_{\mu \nu} = \hat B_{[\mu \nu]}$.
Taking the trace of (\ref{final}) gives
\be
\chi^\sigma \n_{\sigma} \hat \theta = \kappa \hat \theta -
\frac{\hat \theta^2}{d-2} - \hat \sigma_{\mu \nu} \hat \sigma^{\mu
\nu} + \hat \omega_{\mu \nu} \hat \omega^{\mu \nu} - R_{\lambda
\sigma} \chi^{\lambda} \chi^{\s}, \label{ray2}
\ee
while taking the antisymmetric part gives
\be
\chi^{\sigma} \n_{\sigma} \hat \omega_{\mu \nu} = \kappa \hat
\omega_{\mu \nu} - \frac{2}{d-2} \theta \hat \omega_{\mu \nu} - 2
\hat \s^{\s}_{[\nu} \omega_{\mu]\s}\,.
 \label{ray3}
\ee
Finally, the traceless symmetric part is
\be
\chi^{\sigma} \n_{\sigma} \hat \sigma_{\mu \nu} = \kappa \hat
\sigma_{\mu \nu} - \frac{2 \hat \theta}{d-2} \hat \sigma_{\mu \nu}-
\hat \s_{\mu \s}\hat \s^{\s}{}_{\nu} - \hat \omega_{\mu \s} \hat
\omega^{\s}{}_{\nu} + \frac{1}{d-2} \left( \hat \s^2 - \hat
\omega^2\right)Q_{\mu \nu}  - Q^{\al}{}_{\mu} Q^{\beta}{}_{\nu}
C_{\alpha \lambda \beta \s} \chi^{\lambda} \chi^{\s}. \label{ray4}
\ee

We now simplify these equations as in \cite{membrane}. First we note
that the null congruence generating a bifurcate Killing horizon is
hypersurface orthogonal, $i.e.,$
 \be
 \hat \omega_{\mu \nu} = 0.
 \ee
Thus, $\hat B_{\mu \nu}$ is a symmetric 2-tensor identical to the
extrinsic curvature of the cut ${\cal C}$ introduced above. As
usual, this tensor may be written
 \be
  \hat B_{\mu \nu} = \frac{1}{2}
 \mathcal{L}_{\chi} Q_{\mu \nu}.
 \ee
By choosing coordinates adapted to the horizon, we can also replace
the Lie derivative $\mathcal{L}_{\chi}$ by an ordinary derivative
with respect to a Killing parameter $v$:
 \be
  \hat B_{\mu \nu} =
\frac{1}{2} \frac{d Q_{\mu \nu}}{d v}.
 \ee
 We then decompose $\hat
B_{\mu \nu}$ into shear and expansion as before to obtain the
`metric evolution equation' \cite{membrane, damour, PT},
 \be
\frac{1}{2} \frac{d Q_{\mu \nu}}{d v} = \hat \sigma_{\mu \nu} +
\frac{\hat \theta}{d-2} Q_{\mu \nu}.
 \label{evol1}
  \ee

We would like to write (\ref{ray2}), (\ref{ray4}) as similar
ordinary differential equations.  Since the expansion is a scalar,
we can simply replace $ \chi^{\sigma} \n_{\sigma}$ with $\frac{d}{d
v} $ in equation (\ref{ray2}). The result is
 \be
  \frac{d \hat \theta
}{d v} = \kappa \hat \theta - \frac{\hat \theta^2}{d-2} - \hat
\sigma_{\mu \nu} \hat \sigma^{\mu \nu} - R_{\lambda \sigma}
\chi^{\lambda} \chi^{\s}.
 \label{evol2}
 \ee
However, when acting on a tensor quantity like the shear, the
derivatives $\mathcal{L}_\chi$ and $\chi^{\sigma} \n_{\sigma}$
differ by `connection terms' \cite{PT}:
 \bea
 \mathcal{L}_{\chi} \hat \s_{\mu \nu} &=& \chi^\s \n_\s \hat \s_{\mu \nu} + \hat \s_{\s \nu} \n_{\mu} \chi^\s + \hat \s_{\mu \s} \n_{\nu} \chi^\s \\
&=& \chi^{\sigma} \n_{\sigma} \hat \s_{\mu \nu} + \hat \s_{\s \nu} \hat B^{\s}{}_{\mu} + \hat \s_{\mu \s} \hat B^{\s}{}_{\nu} \\
&=& \chi^{\sigma} \n_{\sigma} \hat \s_{\mu \nu} + 2  \hat
\sigma_{\mu}{}^{\s} \hat \s_{\s \nu}  + \frac{2\hat \theta}{d-2}
\hat \s_{\mu \nu}.
 \eea
Thus we find
 \be
 \frac{d\hat \sigma_{\mu \nu}}{d v} = \left( \kappa  - \frac{2
\hat \theta}{d-2}\right) \hat \sigma_{\mu \nu}- \hat \s_{\mu \s}\hat
\s^{\s}{}_{\nu}+ \frac{ \hat \s^2}{d-2}Q_{\mu \nu}  + \left( 2 \hat
\s_{\mu \s} + \frac{2 \hat \theta}{d-2} Q_{\mu \s}\right) \hat
\s^{\s}{}_{\nu} -  Q^{\al}{}_{\mu} Q^{\beta}{}_{\nu} C_{\alpha
\lambda \beta\s} \chi^{\lambda} \chi^{\s}.
 \label{evol3}
  \ee

The `focusing equation' (\ref{evol2}) and the `tidal force equation'
(\ref{evol3}) are the key results of this appendix.   Note that, for
$d=4$, equations (\ref{evol2}) and (\ref{evol3}) can be simplified
because the indices appearing in these equations run only over two
dimensions. In this case we can use the identities
 \be
  \hat \s^{\s}{}_{[\mu}\hat
\omega_{\nu]\s} = 0, \ \ \ \hat \s_{\mu \s} \hat \s^{\s}{}_{\nu} =
\frac{1}{2} \hat \s^2 Q_{\mu \nu} \ \ \ \hat \omega_{\mu \s} \hat
\omega^{\s}{}_{\nu} = - \frac{1}{2} \hat \omega^2 Q_{\mu \nu},
 \ee
 after which our results reproduce those of
\cite{membrane, PT, damour, Suen}.  The equations simplify even
further for $d=3$, where the shear, twist, and Weyl tensor vanish
identically.

\end{document}